\documentclass[a4paper,11pt,draftcls,peerreview,onecolumn]{IEEEtran}

\usepackage{algorithm}
\usepackage[noend]{algorithmic}
\usepackage{algorithmic}

\usepackage{amssymb}
\usepackage[dvips]{graphicx}

\newcommand{\bea}{\begin{eqnarray}}
\newcommand{\eea}{\end{eqnarray}}
\newcommand{\beas}{\begin{eqnarray*}}
\newcommand{\eeas}{\end{eqnarray*}}
\newcommand{\bd}{\begin{displaymath}}
\newcommand{\ed}{\end{displaymath}}
\newcommand{\be}{\begin{equation}}
\newcommand{\ee}{\end{equation}}

\newcommand{\el}{\end{flushleft}}
\newcommand{\bl}{\begin{flushleft}}
\newcommand{\bc}{\begin{center}}
\newcommand{\ec}{\end{center}}
\newcommand{\remove}[1]{}

\newtheorem{theorem}{Theorem}

\newtheorem{lemma}{Lemma}
\newtheorem{defi}{Definition}

\title{Power Optimal Scheduling for Guaranteed Throughput in Multi-access Fading Channels}
\author{Prasanna Chaporkar, Kimmo Kansanen, Ralf R. M\"uller\\
Institute of Electronics and Telecommunications\\
Norwegian University of Science and Technology\\
Trondheim, Norway}
\date{}

\begin{document}
\maketitle
\begin{abstract}
A power optimal scheduling algorithm that guarantees desired throughput
and bounded delay to each user
is developed for fading multi-access multi-band systems.
The optimization is over the joint space of all rate allocation
and coding strategies.
The proposed scheduling assigns rates on each band based only on the current
system state, and subsequently uses optimal multi-user signaling to
achieve these rates.
The scheduling is computationally simple, and hence scalable.
Due to uplink-downlink duality, 
all the results extend in straightforward fashion to the broadcast channels.
\end{abstract}
\newpage
\section{Introduction}
We consider a multi-access fading channel with $N$ users and a single access point. 
Each user $k$ requires certain long term rate ({\it throughput}) 
guarantee $a_k$. {\it Our aim is to design
a scheduling strategy that arbitrates, in every slot, the instantaneous 
rate assignment to each user and coding strategy to realize the
assigned rates depending on the current fading states so that
the throughput requirement for each user is fulfilled and
the total power expenditure is minimized}.

In their seminal work, Tse and Hanly have characterized 
so called {\it throughput capacity} and {\it delay-limited capacity} 
of the multi-access fading channel with Gaussian noise \cite{TH98,HT98}. 
The throughput
capacity region quantifies the achievable rate region with average power
constraint for ergodic fading. 
For the delay limited capacity, each user must be given the required rate
irrespective of its fading states in every slot (strict delay of one slot). 
The aim here is to obtain a coding and
power allocation scheme to minimize the energy.

The notion of throughput capacity leads to schemes that take advantage of
users' differential channel qualities.
Specifically, it is
known that the sum throughput in the system is maximized 
by letting only one user
with the best channel transmit. Schemes that take current channel
states into account while making scheduling decisions are referred to as 
{\it ``Opportunistic Scheduling''} and may result in unfair rate 
allocation if the fading statistics are not symmetric which is typical
in wireless systems.
To alleviate this limitation, several opportunistic scheduling schemes with
fairness constraints have been designed \cite{BBGPSV00,T01}. Among them, 
Proportional Fair Scheduling (PFS) has many desirable properties including
provable fairness guarantees and suitability for on-line implementation, i.e.,
without prior knowledge of channel statistics \cite{VTL02}.
But, PFS does not guarantee the required throughput to users.

Unlike opportunistic scheduling schemes, the delay-limited schemes 
guarantee the
required throughput to every user. Specifically,
super-position encoding and successive decoding is shown to
minimize power for achieving the required throughputs \cite{HT98}.
But, the minimization is achieved under an additional constraint that the
required rate should be provided to each user in each slot irrespective
of its channel state. Thus, these schemes can not benefit from
users' channel variability over time. Recently, we have shown that the
significant power saving can be achieved by exploiting a small delay tolerance
of the application \cite{CKM_rawnet_techreport}. 
In absence of a specific delay constraint, the proposed scheme is
shown to minimize power while guaranteeing the desired throughput 
and bounded delay for each user. The optimality result has been shown in
asymptotic case, i.e., as the number of users go to infinity 
\cite{CKM_rawnet_techreport}.
Optimality for the finite users case has remained open.

For finite users case, \cite{CS06,N05,S05} have found 
back-pressure based scheduling strategies to
minimize the energy consumption in the wireless system with ergodic fading 
while providing the required throughputs and bounded delays to the users. 
These schemes assume that the coding strategy is predefined and 
for the given coding strategy determine the rate to be 
provided in each slot by solving
an optimization problem.
The optimization problem may be non-linear
depending on coding/signaling strategy used, and hence may become
computationally expensive in practice.

Here, we consider the finite users case, 
and propose a computationally simple power optimal scheme that
provides the required throughputs and bounded delays to the users.
The optimization is over the joint space of coding and rate allocations.
Specifically, the proposed optimal policy is 
back-pressure based policy like that in
\cite{CS06,N05,S05}, and employees super-position encoding and
successive decoding in each slot. 
The proposed policy arbitrates scheduling based only on
the users' current backlogs and the channel states. 
In spite of using this limited information,
it is shown to be optimal even in the class of
{\it offline} policies that take into account the channel states and
arrivals in past, present and even the future slots.
One of the main challenges in execution of the proposed policy
is that the optimal rate allocation can only be obtained by solving
a convex optimization in every slot. But, we obtain a computationally
simple algorithm that exploits the problem structure and solves the
optimization.
All the results extend in straightforward fashion to the broadcast case
because of uplink-downlink duality \cite{JVG04}.

The paper is arranged as follows.
In Section~\ref{sec:system_model}, we present our system model.
In Section~\ref{sec:background}, we present some known results that we
use. In Section~\ref{sec:optimal_policy}, we propose our optimal policy and
prove its optimality. In Section~\ref{sec:conclusion}, we conclude.

\section{System Model}
\label{sec:system_model}
We consider a multi-access channel with $N$ users.
Time is slotted. For each user $k$, let $\{A_k(t)\}_{t \ge 1}$
denote the random process of arrivals, i.e., $A_k(t)$ denote the
arrivals for $k$ in slot $t$.
We assume that $\vec{A}(t) = [A_1(t) \ \cdots \ A_N(t)]$ are the
independent and identically distributed (i.i.d.) random vectors across
the slots. Moreover, let $a_k = \mathbb{E}[A_k(t)]$. Alternatively,
$a_k$ denotes the throughput requirement of user $k$. 
We assume that $a_k < \infty$ for every $k$. 
The arrivals for each user $k$ are queued in
the infinite capacity buffer. We denote by
$\vec{Q}(t) = [Q_1(t) \ \cdots \ Q_N(t)]$, where $Q_k(t)$ is the backlog
or queue length for user $k$ in slot $t$, 
i.e., $Q_k(t)$ is the difference between the total arrivals minus the
total departures until time $t$.

Now, we describe our channel model. 
We assume multi-band system. Specifically, we assume that
there are $M$ non-interfering bands available for communication. Let
$\vec{d}_k(t) = [d_{k,1}(t) \ \cdots \ d_{k,M}(t)]$ denote
the vector of channel gains for user $k$ in slot $t$ on each of the bands.
Thus, if $E_{k,m}(t)$ denotes the transmit energy 
per symbol for user $k$ on sub-band
$m$ in slot $t$, then the received energy on the sub-band is given by
$d_{k,m}(t)E_{k,m}(t)$. 
We assume that $\{\vec{d}_k(t): k=1,\ldots,n\}_{t \ge 1}$ is a positive
recurrent finite state Markov process.
Note that this assumption is not restrictive as 
correlated Rician and Rayleigh fading channels can be modeled 
reasonably well using
a finite state Markov process \cite{ZK99,PFL04}.
Let $N_0$ denote the noise power spectral efficiency.

Let $R_{k,m}(t)$ denote the service rate for user $k$ on sub-band $m$
in slot $t$. Then, for every $k$, 
the queue length dynamics is characterized by
\bd
Q_k(t+1) = \max\left\{Q_k(t)+A_k(t) - \sum_{m=1}^M R_{k,m}(t),0\right\}.
\ed
Clearly, $R_{k,m}(t)$ depends on the channel gains, 
transmit energies and the coding strategy used. We consider the space of coding
strategies such that the rates achieved on sub-band $m$ is independent
of the rates on the other sub-bands.
Alternatively, communications on various sub-bands are independent.
Note that the communication on the same sub-band for various users
may not be independent.

\begin{defi}[Scheduling Strategy]
A scheduling policy $\Delta$ arbitrates the rate allocation $R_{k,m}(t)$ and
coding strategy for every user $k$ and sub-band $m$ in every slot $t$.
\end{defi}

This class includes {\it offline} policies that decide their rate
allocation and coding based on the knowledge of
arrivals and channel states in each past, present and even future slots.

We assume that $\vec{Q}(t)$ and $\vec{d}_k(t)$ for every $k$ is
known and a scheduling policy can utilize this knowledge 
in its decision process. In case of a possible ambiguity, we use 
superscript $\Delta$ to indicate the dependence of various terms on 
$\Delta$, e.g., $R_{k,m}^{\Delta}(t)$ and
$E_{k,m}^{\Delta}(t)$ will denote the rate and transmit energy respectively
for user $k$ in sub-band $m$ in slot $t$ under $\Delta$.

\begin{defi}[Stability]
\label{defi:stability}
The multi-access system is said to be stable if the mean queue length 
in every slot $t$ for every
user $k$ is upper bounded by a number that is independent on $t$, i.e.,
 $\sup_{t \ge 1}\{\mathbb{E}[Q_k(t)]\} < \infty$
for every $k$.
A scheduling policy that stabilizes the system is called stable
scheduling policy.
\end{defi}

Note that every stable scheduling policy guarantees the required
throughput $a_k$ to every user $k$, and in addition, guarantees
bounded delay for the arrivals.

\begin{defi}[Power Efficiency]
\label{defi:power_efficiency}
The power efficiency of scheduling policy $\Delta$ is defined as
\bd
P^{\Delta} = \lim\sup_{\!\!\!\!\!\!\!\!\!\!T \to\infty} \frac{1}{T}\sum_{t=1}^T\sum_{k=1}^N\sum_{m=1}^M E_{k,m}^{\Delta}(t).
\ed
\end{defi}
\begin{defi}[Optimality]
\label{defi:epsilon}
A stable policy $\Delta$ is said to be optimal if with probability (w.p.) 1 it
attains the smallest power efficiency among all the stable policies.
\end{defi}

Let $P_{\mathrm{min}}(\mathcal{C})$ be the infimum of the power efficiencies
of all the stable policies in a class $\mathcal{C}$ of scheduling 
policies. If $\mathcal{C}$ does not contain any stable
policy, then $P_{\mathrm{min}}(\mathcal{C})$ is defined to be $\infty$.
Furthermore, let
$P_{\mathrm{min}}$ denote the optimal power efficiency, i.e., $P_{\mathrm{min}}= P_{\mathrm{min}}(\mathcal{C})$ where  $\mathcal{C}$ is the set
of all policies.

\begin{defi}[$\epsilon$-optimality]
\label{defi:epsilon_optimality}
A scheduling policy $\Delta$ is said to be $\epsilon$-optimal in
class $\mathcal{C}$ of scheduling policies if it is stable
and $P^{\Delta} \le P_{\mathrm{min}}(\mathcal{C}) + \epsilon$ w.p.~1.
Moreover, $\Delta$ is said to be $\epsilon$-optimal if it is stable 
and $P^{\Delta} \le P_{\mathrm{min}} + \epsilon$ w.p.~1.
\end{defi}

\section{Background}
\label{sec:background}
We present the following known results for the sake of completeness.
To be consistent, we state these results in the notation introduced here.

Fix a sequence of coding strategies in every slot and let
$\mathcal{C}$ denote the class of scheduling policies that
use this fixed sequence of the coding strategies.
Also, let $\Delta_1(V) \in \mathcal{C}$ denote
a parametrized scheduling policy that assigns the rates by solving the 
optimization problem

\noindent
\textbf{Minimize:} $\sum_{m=1}^M \left[\sum_{k=1}^N V E_{k,m}(t) - \sum_{k=1}^N Q_k(t) R_{k,m}(t)\right]$

\noindent
\textbf{Subject to:} $R_{k,m}(t) \ge 0$ for every $k$ and $m$,

\noindent
where $V$ is a fixed constant. Then, the following are the performance
guarantees for $\Delta_1(V)$.

\begin{theorem}[Result from \cite{CS06,N05,S05}]
\label{thm:known1}
For every $\epsilon > 0$, there exists $\widehat{V} > 0$ such that
for every $V > \widehat{V}$, $\Delta_1(V)$ is $\epsilon$-optimal
in $\mathcal{C}$.
\end{theorem}

We present the intuition for the result.
Consider a case when $Q_k(t)$ is much smaller than $V$. Broadly,
it implies that the user was receiving the desired rate in the past.
Thus, $\Delta_1(V)$ provides positive rate to the user only if the 
corresponding energy cost is much smaller, i.e., when the
user's channel gain is large. On the contrary, 
if $Q_k(t)$ is much larger than $V$, then it implies that the user
was not receiving the desired rate and also that the 
user's average channel gain is small.
Thus, $\Delta_1(V)$ provides positive rate to the user even when 
the user has, potentially, small channel gain in order to preserve
stability. Alternatively, the current queue length represents the history 
of the rate provided to the user and its channel quality.
Thus, $\Delta_1(V)$ estimates users' desired throughput and channel quality
using the current queue length, and then invests just enough power 
to maintain stability.

Given coding strategies, Theorem~\ref{thm:known1} provides a way to
obtain $\epsilon$-optimal policies. 
Thus, it remains to determine how optimal
coding strategy can be obtained in every slot. The following theorem
provides useful guidelines in this direction.

\begin{theorem}[Results in \cite{HT98}]
\label{thm:known2}
For a given rate assignment $R_1,\ldots,R_N$ 
and channel states $d_1,\ldots,d_N$ the total
sum energy $\sum_{k=1}^N E_k$ required to realize the rates
is minimized by super-position coding and successive decoding.
Moreover, for optimal signaling, the successive decoding order depends only
on channel gains, but not on the rate assignment.

Let $\vec{\pi}$ denote the permutation that sorts the gains in the
increasing order, i.e., $d_{\pi_1} \le d_{\pi_2} \le \cdots \le d_{\pi_N}$.
Then, the required transmit energy per symbol for user $\pi_k$ is
given by
\be
\label{equa:energy_os}
E_{\pi_k} = \frac{N_0}{d_{\pi_k}}\left[e^{R_{\pi_k}} - 1\right] e^{\sum_{i < k} R_{\pi_i}}.
\ee
\end{theorem}


\section{$\epsilon$-optimal Scheduling Policy}
\label{sec:optimal_policy}
Let us define the following function for a fixed constant $V$.
\bd
F(\vec{R}_m(t)) \stackrel{\mathrm{def}}{=} \sum_{k=1}^N \frac{V N_0}{d_{\pi_k^m,m}(t)}\left[e^{R_{\pi_k^m,m}(t)} - 1\right] e^{\sum_{i < k} R_{{\pi_i^m,m}}(t)} - \sum_{k=1}^N Q_{\pi_k^m}(t) R_{\pi_k^m,m}(t),
\ed
where $\vec{\pi}^m$ is a permutation that sorts the gains on
sub-band $m$ in the increasing order. Now, let us consider a 
parametrized scheduling policy $\Delta^*(V)$ that assigns in every slot 
the rates $R_{k,m}(t)$ that solve 

\textbf{Optimization (O1) - Minimize:} $\sum_{m=1}^M F(\vec{R}_m(t))$

\textbf{Subject to:} $R_{k,m}(t) \ge 0$ for every $k$ and $m$,

\noindent
and then
achieves the rates using super-position coding and successive decoding
on each sub-band separately. Clearly, $R^{\Delta^*(V)}_{k,m}(t) = 0$ 
for every $m$, if $Q_k(t) = 0$. We show the following
optimality result for $\Delta^*(V)$.

\begin{theorem}
\label{thm:optimal_policy}
For every $\epsilon > 0$, there exists $\widehat{V} > 0$ such that
for every $V > \widehat{V}$, $\Delta^*(V)$ is $\epsilon$-optimal.
\end{theorem}
\proof
Let $\mathcal{C}^*$ denote the class of scheduling policies that use
super-position coding and successive decoding in every slot.
Then, we show that $P_{\min} = P_{\min}(\mathcal{C}^*)$.

Let $\Delta_1$ denote any stable policy. Now, we construct
$\Delta_2 \in \mathcal{C}^*$ as follows.
For every $k$, $m$ and $t$ choose 
$R_{k,m}^{\Delta_2}(t) = R_{k,m}^{\Delta_1}(t)$. 
Clearly, $\Delta_2$ is also stable.
Moreover, by Theorem~\ref{thm:known2}, for every $t$
$\sum_{m=1}^M\sum_{k=1}^N E_{k,m}^{\Delta_2}(t) \le \sum_{m=1}^M\sum_{k=1}^N E_{k,m}^{\Delta_1}(t)$.
Thus, by Definition~\ref{defi:power_efficiency}, $P^{\Delta_1} \ge P^{\Delta_2}$.
Since, $\Delta_1$ is an arbitrary stable scheduling policy, we conclude that
$P_{\min} = P_{\min}(\mathcal{C}^*)$.

Now, the result follows from Theorem~\ref{thm:known1} and 
(\ref{equa:energy_os}).
\endproof

Note that Theorem~\ref{thm:optimal_policy} provides a way to minimize 
power while stabilizing the system. The minimization is over the space
of all coding and rate assignment strategies. 
The policy $\Delta^*(V)$ achieves the
optimality by taking into account only the current system state, and
does not require the knowledge of statistics of the arrival and channel
processes a priori. Moreover, optimality holds among the class of off-line
scheduling policies. In spite of these desirable properties, $\Delta^*(V)$
has one major limitation which is that it needs to solve a non-linear
optimization {\bf(O1)} in every slot to obtain the optimal rate assignment.
Solving {\bf(O1)} may be computationally expensive, and thereby limit
the practicality of $\Delta^*(V)$. In the following discussion, we
focus on {\bf(O1)} and derive certain properties of the 
optimal solution and using these propose an algorithm that obtains 
optimal rate allocation with polynomial complexity.

Since the communication on each of the sub-bands is independent, 
to solve {\bf(O1)}, it suffices to solve separately for every $m$

\textbf{Optimization (O2) - Minimize:} $F(\vec{R}_m(t))$

\textbf{Subject to:} $R_{k,m}(t) \ge 0$ for every $k$.

\noindent
Moreover, since the nature of optimization (objective function and constraints)
 is identical for every $m$, an algorithm to solve {\bf(O2)}
for a given $m$ can be utilized for all $m$'s. 
So, we fix $m$ and $t$ and focus on {\bf(O2)}.

In the following discussion, for notational brevity, we omit $m$ and $t$.
Also, without loss of generality, let $\pi_k^m = k$.
With this simplified notation {\bf(O2)} becomes

\textbf{Minimize:} $F(\vec{R}) = \sum_{k=1}^N \frac{V N_0}{d_{k}}\left[e^{R_k} - 1\right] e^{\sum_{i < k} R_{i}} - \sum_{k=1}^N Q_k R_k$

\textbf{Subject to:} $R_k \ge 0$ for every $k$.

Note that {\bf(O2)} is strictly convex (see Appendix~\ref{app:convex}).
This can be verified by checking that the Hessian is positive definite
in the positive half plane \cite{BV04}.
For convex optimization, polynomial complexity algorithms using
the interior point method have been proposed \cite{NN95}.
These algorithms obtain a solution within $\delta > 0$ neighborhood
of the optimal value.
The computational complexity of these algorithms is $O(N^3)$ per
accuracy digit \cite{NN95}.
We, however, propose the $O(N^2)$ complexity algorithm that computes
the exact optimal solution.
\remove{
\footnote{We believe that the computational
complexity of the proposed algorithm can be further improved by
using specialized data structures and the properties of the optimal
solution. But, these computational issues are out of the scope of 
the this paper.}.
}

We start by looking at the Lagrange relaxation of {\bf(O2)}.

\noindent
\textbf{Minimize:} $F(\vec{R},\vec{\lambda}) = \sum_{k=1}^N \frac{V N_0}{d_{k}}\left[e^{R_k} - 1\right] e^{\sum_{i < k} R_{i}} - \sum_{k=1}^N (Q_k+\lambda_k) R_k$,

\noindent
where $\vec{\lambda} = \{\lambda_1,\ldots,\lambda_N\}$ 
are Lagrange multipliers.
Now, for every $k$
\be
\label{eq:lp_partial1}
\frac{\partial F(\vec{R},\vec{\lambda})}{\partial R_k} = \sum_{i=k+1}^N \frac{V N_0}{d_i} \left(e^{R_i} - 1\right) e^{\sum_{u=1}^{i-1} R_u} + \frac{V N_0}{d_k} e^{\sum_{i=1}^k R_i} - (Q_k + \lambda_k).
\ee

\begin{lemma}
\label{lemma:R}
The following relations satisfy $\frac{\partial F(\vec{R},\vec{\lambda})}{\partial R_k} =0$ for every $k$.
\bea
\label{eq:l_Rk}
R_k &=& \log\left(\frac{[(Q_k+\lambda_k) - (Q_{k+1}+\lambda_{k+1})]\left[\frac{1}{d_{k-1}}-\frac{1}{d_{k}}\right]}{[(Q_{k-1}+\lambda_{k-1}) - (Q_{k}+\lambda_{k})]\left[\frac{1}{d_{k}}-\frac{1}{d_{k+1}}\right]}\right) \mbox{\ \ for $k>1$}\\
\label{eq:l_R1}
R_1 &=& \log\left(\frac{(Q_1+\lambda_1) - (Q_{2}+\lambda_{2})}{V N_0 \left[\frac{1}{d_{1}}-\frac{1}{d_{2}}\right] }\right),
\eea
by defining $d_{N+1} = \infty$ and $Q_{N+1} = \lambda_{N+1} = 0$.
\end{lemma}
\proof
We show the required by proving that for every $k$, $\frac{\partial F(\vec{R},\vec{\lambda})}{\partial R_k} =0$ implies 
\be
\label{eq:lp_partial0}
e^{\sum_{u \le k} R_{u}} = \frac{(Q_k+\lambda_k) - (Q_{k+1} + \lambda_{k+1})}{V N_0 \left[\frac{1}{d_k} - \frac{1}{d_{k+1}}\right]}.
\ee
We prove the above using induction on $k$.

As a base case
we show (\ref{eq:lp_partial0}) for $k=N$.
Note that substituting $k=N$ in (\ref{eq:lp_partial1}) and equating it to 0, 
we obtain (\ref{eq:lp_partial0}). 
Thus, (\ref{eq:lp_partial0}) holds for 
$k = N$. Now, for induction, we assume that
(\ref{eq:lp_partial0}) holds for
every $k \ge s+1$ and verify it for $k=s$.

Consider the second term in (\ref{eq:lp_partial1}) with $k=s$.
\beas
\lefteqn{\sum_{i=s+1}^N \frac{V N_0}{d_i} \left(e^{R_i} - 1\right) e^{\sum_{u=1}^{i-1} R_u}} \\
&=& \sum_{i=s+1}^N \frac{V N_0}{d_i} e^{\sum_{u=1}^{i} R_u} - \sum_{i=s+1}^N \frac{V N_0}{d_i} e^{\sum_{u=1}^{i-1} R_u} \\
&=& \sum_{i=s+1}^{N-1} V N_0\left[\frac{1}{d_i}-\frac{1}{d_{i+1}}\right] e^{\sum_{u=1}^{i} R_u} + \frac{V N_0}{d_N} e^{\sum_{u=1}^{N} R_u} - \frac{V N_0}{d_{s+1}} e^{\sum_{u=1}^{s} R_u} \\
&=& (Q_{s+1} + \lambda_{s+1}) - \frac{V N_0}{d_{s+1}} e^{\sum_{u=1}^{s} R_u}.
\eeas
Last equality follows from (\ref{eq:lp_partial0}) and the induction hypothesis.
Now, substituting the above in (\ref{eq:lp_partial1}), we obtain the
desired.

Finally, (\ref{eq:l_Rk}) follows by observing $R_k = \log\left(\frac{e^{\sum_{u=1}^{k}}R_u}{e^{\sum_{u=1}^{k-1}}R_u}\right)$ and (\ref{eq:l_R1}) is obtained directly
from (\ref{eq:lp_partial0}) with $k=1$.
\endproof

\begin{defi}[From \cite{NW99}, pp. 328]
The vectors $\vec{R}^{\prime}$ and $\vec{\lambda}^{\prime}$ are said to satisfy 
Karush-Kuhn-Tucker (KKT) conditions if
they satisfy the following relations.
\bea
\label{eq:KKT1}
\left. \frac{\partial F(\vec{R},\vec{\lambda})}{\partial R_k} \right|_{\vec{R} = \vec{R}^{\prime}} &=& 0 \mbox{\ \ for every $k$} \\
\label{eq:KKT2}
\vec{R}^{\prime} &\ge& \vec{0} \\
\label{eq:KKT3}
\vec{\lambda}^{\prime} &\ge& \vec{0} \\
\label{eq:KKT4}
R_k^{\prime} {\lambda}_k^{\prime} &=& 0 \mbox{\ \ for every $k$}.
\eea
\end{defi}

Since {\bf({O2})} is strictly convex in the feasible region, we conclude the
following \cite{NW99}. 
\begin{enumerate}
\item
The optimal solution is unique.
\item
The rate allocation $\vec{R}^{\prime}$ is optimal {\it iff}
there exists $\vec{\lambda}^{\prime}$ such that 
$\vec{R}^{\prime}$ and $\vec{\lambda}^{\prime}$ satisfy the
KKT conditions. Also, such $\vec{\lambda}^{\prime}$ is unique
since linear independence constraint qualification holds.
\end{enumerate}

In Figure~\ref{fig:opt_algo}, we propose a general procedure for obtaining
a rate allocation $\vec{R}$ and Lagrange multipliers $\vec{\lambda}$ 
that satisfy the KKT conditions
for any given $\vec{Q}$ and $\vec{d}$. 
We first intuitively describe the proposed algorithm and subsequently
prove that the algorithm optimally solves {\bf (O2)}.

\begin{figure}[tb]
\mbox{}\hrulefill
\vspace{.3em}
\\Computation\_of\_Optimal\_Rates($\vec{Q}$,$\vec{d}$) \\
begin
\begin{scriptsize}
\begin{algorithmic}[1]
\STATE Initialize $\mathcal{A} \leftarrow \{1,\ldots,N\}$, $\mathcal{E} \leftarrow \phi$ and $\vec{\lambda} \leftarrow \vec{0}$
\label{line11}
\WHILE {There exists $k \in \mathcal{A}$ such that
\label{line12}
\bea
\label{eq:algo1}
Q_k &<& \frac{(Q_{k-1} + \lambda_{k-1})\left[\frac{1}{d_k}-\frac{1}{d_{k+1}}\right] + (Q_{k+1} + \lambda_{k+1})\left[\frac{1}{d_{k-1}}-\frac{1}{d_{k}}\right]}{\left[\frac{1}{d_{k-1}}-\frac{1}{d_{k+1}}\right]}  \mbox{ \ \ for $k>1$} \\
\label{eq:algo2}
Q_1 &<& V N_0 \left[\frac{1}{d_{1}}-\frac{1}{d_{2}}\right] + (Q_{2}+\lambda_{2})
\eea
}
\STATE $\mathcal{E} \leftarrow \mathcal{E} \cup \{k\}$
\label{line13}
\STATE $\mathcal{A} \leftarrow \mathcal{A} - \{k\}$
\label{line14}
\STATE Update\_Lagrange\_Multipliers($\mathcal{A},\mathcal{E}$)
\label{line15}
\ENDWHILE
\COMMENT {/* Optimal Rate computation */}
\STATE $R_k \leftarrow 0$ for every $k \in \mathcal{E}$ \\
\label{line16}
\STATE Compute $R_k$ for every $k \in \mathcal{A}$ using (\ref{eq:l_Rk}) and (\ref{eq:l_R1}) \\
\label{line17}
\end{algorithmic}
\end{scriptsize}
end\\

Update\_Lagrange\_Multipliers($\mathcal{A},\mathcal{E}$) \\
begin
\begin{scriptsize}
\begin{algorithmic}[1]
\STATE $\lambda_k \leftarrow 0$ for every $k \in \mathcal{E}$ \\
\label{line21}
\IF{$\{1,\ldots, u-1\} \subseteq \mathcal{E}$ and $u \in \mathcal{A}$} 
\label{line22}
\STATE for every $m \in \{1,\ldots,u-1\}$
\label{line23}
\be
\label{eq:algo_lambda_1}
\lambda_m \leftarrow V N_0\left[\frac{1}{d_m}-\frac{1}{d_{u}}\right] + (Q_{u}-Q_m).
\ee
\ENDIF
\IF{$\{v+1,\ldots,u-1\} \subseteq \mathcal{A}$ and $\{v,u\} \subseteq \mathcal{E}$} 
\STATE for every
$m \in \{v+1,\ldots,u-1\}$
\label{line25}
\be
\label{eq:algo_lambda_2}
\lambda_m \leftarrow \frac{Q_v\left[\frac{1}{d_m}-\frac{1}{d_{u}}\right] + Q_u\left[\frac{1}{d_v}-\frac{1}{d_{m}}\right]}{\left[\frac{1}{d_v}-\frac{1}{d_{u}}\right]} - Q_m.
\ee
\ENDIF
\end{algorithmic}
\end{scriptsize}
end\\
\vspace{.3em}
\mbox{}\hrulefill
\caption{\label{fig:opt_algo} Figure shows the pseudo code of an algorithm
that  computes the optimal rate allocation in a given slot}
\end{figure}

The main procedure Computation\_of\_Optimal\_Rates takes current  queue
length vector $\vec{Q}$ and the channel gains $\vec{d}$ as input and
outputs the optimal rate allocation $\vec{R}$.
In this procedure, we define two sets $\mathcal{A}$ and $\mathcal{E}$
that partition the set of all users.
The set $\mathcal{A}$ ($\mathcal{E}$, resp.) denotes the set of active
(inactive, resp.) users. A user $k$ is said to be active if $R_k > 0$, i.e.,
it is served at positive rate; $k$ is inactive otherwise.
 Initially, all the users are assumed to be
active (Line~\ref{line11}). Next, the algorithm iterates and in each iteration
determines an inactive user using (\ref{eq:algo1}) and (\ref{eq:algo2})
(Line~\ref{line12}).
Once the inactive user is determined the sets $\mathcal{A}$ and $\mathcal{E}$
are updated (Lines~\ref{line13} and~\ref{line14}),
and subsequently the Lagrange multipliers are also updated (Line~\ref{line15}).
If no user in $\mathcal{A}$ satisfy (\ref{eq:algo1}) and (\ref{eq:algo2}), 
then the algorithm terminates after computing 
the rate allocation using (\ref{eq:l_Rk}) and
(\ref{eq:l_R1}) (Lines~\ref{line16} and~\ref{line17}). 
This ensures that (\ref{eq:KKT1}) is satisfied 
for all $k \in \mathcal{E}$.
Now, we explain why a user satisfying (\ref{eq:algo1}) or (\ref{eq:algo2})
should be inactive. Note that (\ref{eq:algo1}) and (\ref{eq:algo2}) are
equivalent to $R_k < 0$ in (\ref{eq:l_Rk}) and
(\ref{eq:l_R1}), respectively. Since the assigned rates can only be 
non-negative, we put such a user $k$ in $\mathcal{E}$ and update
corresponding $\lambda_k$ so as to ensure $R_k = 0$.

Now, we briefly explain how the procedure Update\_Lagrange\_Multipliers
computes Lagrange multipliers in each iteration.
Note that for every active user $k$, $\lambda_k$ must be zero
in order to satisfy the KKT condition (\ref{eq:KKT4}).
Thus in the first step, the procedure assigns $\lambda_k = 0$
for every $k \in \mathcal{A}$ (Line~\ref{line21}). 
Next, for every $k \in \mathcal{E}$,
it computes $\lambda_k$ so that $R_k$ in (\ref{eq:l_Rk}) or
(\ref{eq:l_R1}) equals zero (Lines~\ref{line22} to~\ref{line25}). 
This ensures that (\ref{eq:KKT1}),
(\ref{eq:KKT2}) and (\ref{eq:KKT4}) hold for every $k \in \mathcal{A}$.
We need to recompute all the Lagrange multipliers in every iteration 
because the value of $\lambda_k$ is a function of $\lambda_{k-1}$ and
$\lambda_{k+1}$ as can be seen from (\ref{eq:l_Rk}) and
(\ref{eq:l_R1}).

Even though the algorithm is straightforward, mainly, two questions
are unanswered. First, whether $\lambda_k$ is non-negative for
every $k \in \mathcal{E}$. Second, since the $\lambda_k$'s for many
users (not only the recently added user) in $\mathcal{E}$ are updated,
how is it ensured that an inactive user does not become active in the
subsequent iterations. We formally address these questions and prove
the optimality of the proposed algorithm.

For analysis, we introduce the following additional notation.
Let $\vec{R}^*$ and $\vec{\lambda}^*$ denote the rate vector and 
Lagrange multipliers computed by the algorithm at termination. 
Also, let $\mathcal{A}^*$ and $\mathcal{E}^*$ denote the sets
$\mathcal{A}$ and $\mathcal{E}$, respectively, when the algorithm terminates.
Next, we distinguish between the value of $\vec{\lambda}$,
$\mathcal{A}$ and $\mathcal{E}$ 
computed by the algorithms in every iteration. 
Let $\vec{\lambda}^i$, $\mathcal{A}^i$ and $\mathcal{E}^i$ denote
$\vec{\lambda}$, $\mathcal{A}$ and $\mathcal{E}$, respectively,
computed by the algorithm in $i^{\mathrm{th}}$ iteration.
Because of the initialization in Line~\ref{line11} of procedure
Computation\_of\_Optimal\_Rates, 
$\vec{\lambda}^0 = \vec{0}$, $\mathcal{A}^0 = \{1,\ldots,N\}$ 
and $\mathcal{E}^0 = \phi$.
Let the algorithm terminate in $I$ iterations. Then, clearly, 
$I \le N$ and $\vec{\lambda}^I = \vec{\lambda}^*$,
$\mathcal{A}^I = \mathcal{A}^*$ and $\mathcal{E}^I = \mathcal{E}^*$. 
Now, we show the following result.

\begin{lemma}
\label{lemma:KKT}
If $\vec{\lambda}^* \ge \vec{0}$, then $\vec{R}^*$ and $\vec{\lambda}^*$
satisfy the KKT conditions.
\end{lemma}
\proof
Note that for every $k \in \mathcal{A}^*$, $R_k^*$ is computed using
(\ref{eq:l_Rk}) and (\ref{eq:l_R1}). Thus by Lemma~\ref{lemma:R},
clearly, (\ref{eq:KKT1}) is satisfied for every $k \in \mathcal{A}^*$.
Now, we show that (\ref{eq:KKT1}) also holds for every $k \in \mathcal{E}^*$.
Note that $R_k^* = 0$ for every $k \in \mathcal{E}^*$.
Thus, it suffices to show that when the chosen $\vec{\lambda}^*$ 
is substituted in (\ref{eq:l_Rk}) and (\ref{eq:l_R1}) yields 
$R_k^* = 0$ for every $k \in \mathcal{E}^*$. The required can be easily
verified using elementary algebra. Thus (\ref{eq:KKT1}) holds for every $k$.

Now, we show that $\vec{\lambda}^*$ satisfy (\ref{eq:KKT2}).
Since, $R_k^* = 0$ for every $k \in \mathcal{E}^*$, (\ref{eq:KKT2})
clearly holds for every $k \in \mathcal{E}^*$. Now, we show (\ref{eq:KKT2})
for every $k \in \mathcal{A}^*$. We show the required using contradiction.
Let there be $k \in \mathcal{A}^*$ such that $R_k^* < 0$.
But then from (\ref{eq:l_Rk}) and (\ref{eq:l_R1}) it implies that
\beas
Q_k &<& \frac{(Q_{k-1} + \lambda_{k-1}^*)\left[\frac{1}{d_k}-\frac{1}{d_{k+1}}\right]+(Q_{k+1} + \lambda_{k+1}^*)\left[\frac{1}{d_{k-1}}-\frac{1}{d_{k}}\right]}{\left[\frac{1}{d_{k-1}}-\frac{1}{d_{k+1}}\right]}  \mbox{ \ \ if $k>1$} \\
Q_k &<& V N_0 \left[\frac{1}{d_{k}}-\frac{1}{d_{k+1}}\right] + (Q_{k+1}+\lambda_{k+1}^*) \mbox{ \ \ if $k=1$}.
\eeas
Now, from (\ref{eq:algo1}) and (\ref{eq:algo2}), we conclude that the algorithm
will not terminate, but instead add $k$ to $\mathcal{E}$ and continue. 
Thus, no such index exists. 
So, $\vec{\lambda}^*$ and $\vec{R}^*$ satisfy (\ref{eq:KKT2}).

The vectors  $\vec{\lambda}^*$ and $\vec{R}^*$ satisfy (\ref{eq:KKT3}) 
because of the supposition in the lemma. Moreover,
the vectors satisfy (\ref{eq:KKT4}) because $R_k^* = 0$ for
every $k \in \mathcal{E}^*$, while $\lambda_k^* = 0$ for
every $k \in \mathcal{A}^*$.
\endproof

In the following theorem, we show that $\vec{\lambda}^*$ is non-negative.

\begin{theorem}
\label{thm:lambda}
For every $i < I$, $\vec{\lambda}^i \le \vec{\lambda}^{i+1}$.
\end{theorem}

Note that since $\vec{\lambda}^0 = \vec{0}$, Theorem~\ref{thm:lambda}
implies that $\vec{\lambda}^* \ge \vec{0}$.
We prove the above theorem by showing the required in each of
the cases that may be arise in the execution of the algorithm.
The proofs use elementary algebra. 
For better readability, proofs for all
the cases are given in Appendix~\ref{app:thm4}.

Finally, we prove the optimality of the proposed algorithm.
\begin{theorem}
The rate allocation $\vec{R}^*$ is the unique optimal solution of {\bf(O2)}.
\end{theorem}
\proof
The result follows immediately from the strict convexity of {\bf(O2)},
Lemma~\ref{lemma:KKT} and Theorem~\ref{thm:lambda}.
\endproof

\section{Conclusion}
\label{sec:conclusion}
We have considered a multi-access channel with $N$-users. 
We have proposed a parametrized scheduling policy $\Delta^*(V)$
which is $\epsilon$-optimal for every $\epsilon > 0$ for appropriate
choice of the parameter $V$ even among the offline strategies in spite of
considering only the current queue lengths and channel gains in its decision
process. Moreover, the optimization is over the joint space of
coding and rate allocation strategies. The policy $\Delta^*(V)$ needs to
solve a convex optimization in every slot to obtain the optimal rate 
allocation. We have proposed a $O(N^2)$ algorithm that accurately solves the
optimization.
All the results extend in straightforward fashion to broadcast case
because of uplink-downlink duality.


\appendices
\section{Convexity of Optimization {\bf (O2)}}
\label{app:convex}
The second partial derivative of $F(\vec{R})$ is as follows.
\bd
\frac{{\partial}^2F(\vec{R})}{\partial R_k \partial R_j} = \left\{ \begin{array}
{r@{\quad:\quad}l}
\sum_{i=k+1}^N \frac{VN_0}{d_i}\left(e^{R_i} -1\right) e^{\sum_{u=1}^{i-1} R_u} + \frac{VN_0}{d_k} e^{\sum_{i=1}^{k} R_i} & j \le k \\
\frac{VN_0}{d_j} e^{\sum_{u=1}^{j} R_u} + \sum_{i=j+1}^N \frac{VN_0}{d_i}\left(e^{R_i} -1\right) e^{\sum_{u=1}^{i-1} R_u} & j > k.
\end{array} \right.
\ed

Note that for every $\vec{R} \in [0,\infty)^N$, $\frac{{\partial}^2F(\vec{R})}{\partial R_k \partial R_j} >0$ for any $k$ and $j$.
This shows that the Hessian of $F(\vec{R})$ is
positive definite. Also, it is clear that the feasible region
$[0,\infty)^N$ is a convex set. Thus, {\bf (O2)} is an instance of
convex optimization.
\label{app:convexity}
\section{Supporting Lemmas for Proving Theorem~\ref{thm:lambda}}
\label{app:thm4}
\begin{lemma}
\label{lemma:change}
Let index $k$ be added to the set $\mathcal{E}^{i-1}$ in the
$i^{\mathrm{th}}$ iteration. Then for all users $u$ such that
there exists $v \in \mathcal{A}^{i}$ between $k$ and $u$, 
$\lambda_u^{i} = \lambda_u^{i-1}$. 
\end{lemma}
\proof
The proof follows immediately from the procedure
Update\_Lagrange\_Multipliers in Figure~\ref{fig:opt_algo}.
\endproof
\begin{lemma}
\label{lemma:isolated}
Let index $k$ be added to the set $\mathcal{E}^{i-1}$ in the
$i^{\mathrm{th}}$ iteration. Also, let $\{k-1,k+1\} \in \mathcal{A}^{i-1}$.
Then, $\lambda_n^{i} - \lambda_n^{i-1} \ge 0$ for every $n$.
\end{lemma}
\proof
Since index $k$ is added to $\mathcal{E}^{i-1}$ in the
$i^{\mathrm{th}}$ iteration, we know the following.
First, $\lambda_k^{i-1} = 0$. Second, from (\ref{eq:algo1})
\be
\label{eq:lip_1}
Q_k < \frac{(Q_{k-1} + \lambda_{k-1}^{i-1})\left[\frac{1}{d_k} - \frac{1}{d_{k+1}}\right] + (Q_{k+1} + \lambda_{k+1}^{i-1}) )\left[\frac{1}{d_{k-1}} - \frac{1}{d_{k}}\right]}{\left[\frac{1}{d_{k-1}} - \frac{1}{d_{k+1}}\right]}.
\ee
Note that $\lambda_{k-1}^{i-1} = \lambda_{k+1}^{i-1} =0$. Thus the result
follows from (\ref{eq:algo_lambda_2}), (\ref{eq:lip_1}) and Lemma~\ref{lemma:change}.
\endproof

\begin{lemma}
\label{lemma:l3}
Let index 1 be added to the set $\mathcal{E}^{i-1}$ in the
$i^{\mathrm{th}}$ iteration. Also, let 
$\{2,\ldots,m-1\} \subseteq \mathcal{E}^{i-1}$ 
and $m \not\in \mathcal{E}^{i-1}$.
Then, $\lambda_n^{i} - \lambda_n^{i-1} \ge 0$ for every $n$.
\end{lemma}
\proof
Since index 1 is added to $\mathcal{E}^{i-1}$ in the
$i^{\mathrm{th}}$ iteration, we know the following.
First, $\lambda_1^{i-1} = 0$. Second, from (\ref{eq:algo2})
\be
\label{eq:l3p_1}
Q_1 < V N_0 \left[\frac{1}{d_1} - \frac{1}{d_2}\right] + (Q_2 + \lambda_2^{i-1}).
\ee

Moreover, since $\{2,\ldots,m-1\} \subseteq \mathcal{E}^{i-1}$
and $m \not\in \mathcal{E}^{i-1}$, we also know that
for every $n \in \{2,\ldots,m-1\}$,
\be
\label{eq:l3p_2}
\lambda_n^{i-1} = \frac{Q_1 \left[\frac{1}{d_n} - \frac{1}{d_m}\right] + Q_m\left[\frac{1}{d_1} - \frac{1}{d_n}\right]}{\left[\frac{1}{d_1} - \frac{1}{d_m}\right]} - Q_n.
\ee
Now, substituting $\lambda_2^{i-1}$ from (\ref{eq:l3p_2}) in (\ref{eq:l3p_1}),
we obtain
\bea
Q_1 &<& V N_0 \left[\frac{1}{d_1} - \frac{1}{d_2}\right] + \frac{Q_1 \left[\frac{1}{d_2} - \frac{1}{d_m}\right] + Q_m\left[\frac{1}{d_1} - \frac{1}{d_2}\right]}{\left[\frac{1}{d_1} - \frac{1}{d_m}\right]} \nonumber \\
\label{eq:l3p_3}
\Longrightarrow Q_1 &<& V N_0 \left[\frac{1}{d_1} - \frac{1}{d_m}\right] + Q_m. 
\eea
Now, note that from (\ref{eq:algo_lambda_1}), for every $n = 1,\ldots,m-1$
\be
\label{eq:l3p_4}
\lambda_n^{i} = V N_0 \left[\frac{1}{d_n} - \frac{1}{d_m}\right] + (Q_m-Q_n).
\ee
From (\ref{eq:l3p_3}), clearly, $\lambda_1^{i} > 0$.
Now, from (\ref{eq:l3p_2}) and (\ref{eq:l3p_4}), it follows that for
every $n = 2,\ldots,m-1$
\beas
\lambda_n^{i} - \lambda_n^{i-1} &=& V N_0 \left[\frac{1}{d_n} - \frac{1}{d_m}\right] + (Q_m-Q_n) - \frac{Q_1 \left[\frac{1}{d_n} - \frac{1}{d_m}\right] + Q_m\left[\frac{1}{d_1} - \frac{1}{d_n}\right]}{\left[\frac{1}{d_1} - \frac{1}{d_m}\right]} + Q_n \\
&=& \frac{\left[\frac{1}{d_n} - \frac{1}{d_m}\right]}{\left[\frac{1}{d_1} - \frac{1}{d_m}\right]} \lambda_1^i \ge 0.
\eeas
The last inequality follows from the fact that $d_k \le d_{k+1}$ for every $k$
and $ \lambda_1^i > 0$.
Furthermore, by  Lemma~\ref{lemma:change}, for $n\not\in\{1,\ldots,m-1\}$, $\lambda_n^{i-1} = \lambda_n^{i}$.
Thus, the result follows.
\endproof

\begin{lemma}
\label{lemma:l4}
Let index $k>1$ be added to the set $\mathcal{E}^{i-1}$ in the
$i^{\mathrm{th}}$ iteration. Also, let 
$\{v+1,\ldots,k-1\} \subseteq \mathcal{E}^{i-1}$ 
and $\{v,k+1\} \subseteq \mathcal{A}^{i}$.
Then, $\lambda_n^{i} - \lambda_n^{i-1} \ge 0$ for every $n$.
\end{lemma}
\proof
Since index $k$ is added to $\mathcal{E}^{i-1}$ in the
$i^{\mathrm{th}}$ iteration, we know the following.
First, $\lambda_k^{i-1} = 0$. Second, from (\ref{eq:algo1})
\be
\label{eq:l4p_1}
Q_k < \frac{(Q_{k-1} + \lambda_{k-1}^{i-1})\left[\frac{1}{d_k} - \frac{1}{d_{k+1}}\right] + (Q_{k+1} + \lambda_{k+1}^{i-1}) )\left[\frac{1}{d_{k-1}} - \frac{1}{d_{k}}\right]}{\left[\frac{1}{d_{k-1}} - \frac{1}{d_{k+1}}\right]}.
\ee

Moreover, since $\{v+1,\ldots,k-1\} \subseteq \mathcal{E}^{i-1}$
and $\{v,k,k+1\} \subseteq \mathcal{A}^{i-1}$, we also know that
for every $n \in \{v+1,\ldots,k-1\}$,
\bea
\label{eq:l4p_2}
\lambda_n^{i-1} = \frac{Q_v \left[\frac{1}{d_n} - \frac{1}{d_k}\right] + Q_k\left[\frac{1}{d_v} - \frac{1}{d_n}\right]}{\left[\frac{1}{d_v} - \frac{1}{d_k}\right]} - Q_n.
\eea
Now, substituting $\lambda_{k+1}^{i-1} = 0$ and $\lambda_{k-1}^{i-1}$ from 
(\ref{eq:l4p_2}) in (\ref{eq:l4p_1}),
we obtain
\bea
Q_k &<& \frac{\frac{Q_v \left[\frac{1}{d_{k-1}} - \frac{1}{d_k}\right] + Q_k\left[\frac{1}{d_v} - \frac{1}{d_{k-1}}\right]}{\left[\frac{1}{d_v} - \frac{1}{d_k}\right]}\left[\frac{1}{d_k} - \frac{1}{d_{k+1}}\right] + (Q_{k+1} + \lambda_{k+1}^{i-1}) )\left[\frac{1}{d_{k-1}} - \frac{1}{d_{k}}\right]}{\left[\frac{1}{d_{k-1}} - \frac{1}{d_{k+1}}\right]} \nonumber \\
\label{eq:l4p_3}
\Longrightarrow Q_k &<& \frac{Q_v \left[\frac{1}{d_{k}} - \frac{1}{d_{k+1}}\right] + Q_{k+1}\left[\frac{1}{d_v} - \frac{1}{d_k}\right]}{\left[\frac{1}{d_v} - \frac{1}{d_{k+1}}\right]}. 
\eea
Now, note that from (\ref{eq:algo_lambda_2}), for every $n = v+1,\ldots,k$
\be
\label{eq:l4p_4}
\lambda_n^{i} =  \frac{Q_v \left[\frac{1}{d_{n}} - \frac{1}{d_{k+1}}\right] + Q_{k+1}\left[\frac{1}{d_v} - \frac{1}{d_n}\right]}{\left[\frac{1}{d_v} - \frac{1}{d_{k+1}}\right]}-Q_n.
\ee
From (\ref{eq:l4p_3}) and (\ref{eq:l4p_4}), clearly, $\lambda_k^{i} > 0$.
Now, from (\ref{eq:l4p_2}) and (\ref{eq:l4p_4}), it follows that for
every $n = v+1,\ldots,k-1$
\beas
\lambda_n^{i} - \lambda_n^{i-1} &=& \frac{Q_v \left[\frac{1}{d_{n}} - \frac{1}{d_{k+1}}\right] + Q_{k+1}\left[\frac{1}{d_v} - \frac{1}{d_n}\right]}{\left[\frac{1}{d_v} - \frac{1}{d_{k+1}}\right]} - \frac{Q_v \left[\frac{1}{d_n} - \frac{1}{d_k}\right] + Q_k\left[\frac{1}{d_v} - \frac{1}{d_n}\right]}{\left[\frac{1}{d_v} - \frac{1}{d_k}\right]} \\
&=& \frac{\left[\frac{1}{d_v} - \frac{1}{d_n}\right]}{\left[\frac{1}{d_v} - \frac{1}{d_k}\right]} \lambda_k^i  \ge 0.
\eeas
The last inequality follows from the fact that $d_n \le d_{n+1}$ for every $n$
and $ \lambda_k^i > 0$.
Furthermore, by  Lemma~\ref{lemma:change}, for $n\not\in\{v+1,\ldots,k\}$, $\lambda_n^{i-1} = \lambda_n^{i}$.
Thus, the result follows.
\endproof

\begin{lemma}
\label{lemma:l5}
Let index $k>1$ be added to the set $\mathcal{E}^{i-1}$ in the
$i^{\mathrm{th}}$ iteration. Also, let 
$\{1,\ldots,k-1\} \subseteq \mathcal{E}^{i-1}$ 
and $k+1 \in \mathcal{A}^{i}$.
Then, $\lambda_n^{i} - \lambda_n^{i-1} \ge 0$ for every $n$.
\end{lemma}
\proof
Since index $k$ is added to $\mathcal{E}^{i-1}$ in the
$i^{\mathrm{th}}$ iteration, we know the following.
First, $\lambda_k^{i-1} = 0$. Second, from (\ref{eq:algo1})
\be
\label{eq:l5p_1}
Q_k < \frac{(Q_{k-1} + \lambda_{k-1}^{i-1})\left[\frac{1}{d_k} - \frac{1}{d_{k+1}}\right] + (Q_{k+1} + \lambda_{k+1}^{i-1}) )\left[\frac{1}{d_{k-1}} - \frac{1}{d_{k}}\right]}{\left[\frac{1}{d_{k-1}} - \frac{1}{d_{k+1}}\right]}.
\ee

Moreover, since $\{1,\ldots,k-1\} \subseteq \mathcal{E}^{i-1}$
and $k \in \mathcal{A}^{i-1}$, we also know that
for every $n \in \{1,\ldots,k-1\}$,
\be
\label{eq:l5p_2}
\lambda_n^{i-1} = V N_0 \left[\frac{1}{d_n}-\frac{1}{d_k}\right] +(Q_k - Q_n).
\ee
Now, substituting $\lambda_{k+1}^{i-1} = 0$ and $\lambda_{k-1}^{i-1}$ from 
(\ref{eq:l5p_2}) in (\ref{eq:l5p_1}),
we obtain
\bea
Q_k &<& \frac{\left(V N_0 \left[\frac{1}{d_{k-1}}-\frac{1}{d_k}\right] + Q_k\right) \left[\frac{1}{d_k} - \frac{1}{d_{k+1}}\right] + Q_{k+1} \left[\frac{1}{d_{k-1}} - \frac{1}{d_{k}}\right]}{\left[\frac{1}{d_{k-1}} - \frac{1}{d_{k+1}}\right]} \\
\label{eq:l5p_3}
\Longrightarrow Q_k &<&  V N_0 \left[\frac{1}{d_k}-\frac{1}{d_{k+1}}\right] +Q_{k+1}.
\eea
Now, note that from (\ref{eq:algo_lambda_2}), for every $n = 1,\ldots,k$
\be
\label{eq:l5p_4}
\lambda_n^{i} =   V N_0 \left[\frac{1}{d_n}-\frac{1}{d_{k+1}}\right] + (Q_{k+1}-Q_n).
\ee
From (\ref{eq:l5p_3}) and (\ref{eq:l5p_4}), clearly, $\lambda_k^{i} > 0$.
Now, from (\ref{eq:l5p_2}) and (\ref{eq:l5p_4}), it follows that for
every $n = 1,\ldots,k-1$
\beas
\lambda_n^{i} - \lambda_n^{i-1} &=& V N_0 \left[\frac{1}{d_k}-\frac{1}{d_{k+1}}\right] + (Q_{k+1}-Q_k) \\
&=& \lambda_k^i  \ge 0.
\eeas
Furthermore, by  Lemma~\ref{lemma:change}, for $n\not\in\{1,\ldots,k\}$, $\lambda_n^{i-1} = \lambda_n^{i}$.
Thus, the result follows.
\endproof

\begin{lemma}
\label{lemma:l6}
Let index $k>1$ be added to the set $\mathcal{E}^{i-1}$ in the
$i^{\mathrm{th}}$ iteration. Also, let 
$\{k+1,\ldots,u-1\} \subseteq \mathcal{E}^{i-1}$ 
and $\{k-1,u\} \subseteq \mathcal{A}^{i}$.
Then, $\lambda_n^{i} - \lambda_n^{i-1} \ge 0$ for every $n$.
\end{lemma}
\proof
Since index $k$ is added to $\mathcal{E}^{i-1}$ in the
$i^{\mathrm{th}}$ iteration, we know the following.
First, $\lambda_k^{i-1} = 0$. Second, from (\ref{eq:algo1})
\be
\label{eq:l6p_1}
Q_k < \frac{(Q_{k-1} + \lambda_{k-1}^{i-1})\left[\frac{1}{d_k} - \frac{1}{d_{k+1}}\right] + (Q_{k+1} + \lambda_{k+1}^{i-1}) )\left[\frac{1}{d_{k-1}} - \frac{1}{d_{k}}\right]}{\left[\frac{1}{d_{k-1}} - \frac{1}{d_{k+1}}\right]}.
\ee

Moreover, since $\{k+1,\ldots,u-1\} \subseteq \mathcal{E}^{i-1}$
and $\{k-1,k,u\} \subseteq \mathcal{A}^{i-1}$, we also know that
for every $n \in \{k+1,\ldots,u-1\}$,
\bea
\label{eq:l6p_2}
\lambda_n^{i-1} = \frac{Q_k \left[\frac{1}{d_n} - \frac{1}{d_u}\right] + Q_u\left[\frac{1}{d_k} - \frac{1}{d_n}\right]}{\left[\frac{1}{d_k} - \frac{1}{d_u}\right]} - Q_n.
\eea
Now, substituting $\lambda_{k-1}^{i-1} = 0$ and $\lambda_{k+1}^{i-1}$ from 
(\ref{eq:l6p_2}) in (\ref{eq:l6p_1}),
we obtain
\bea
Q_k &<& \frac{Q_{k-1}\left[\frac{1}{d_{k}} - \frac{1}{d_{k+1}}\right] + \frac{Q_k \left[\frac{1}{d_{k+1}} - \frac{1}{d_u}\right] + Q_u\left[\frac{1}{d_k} - \frac{1}{d_{k+1}}\right]}{\left[\frac{1}{d_k} - \frac{1}{d_u}\right]}\left[\frac{1}{d_{k-1}} - \frac{1}{d_{k}}\right]}{\left[\frac{1}{d_{k-1}} - \frac{1}{d_{k+1}}\right]} \nonumber \\ 
\label{eq:l6p_3}
\Longrightarrow Q_k &<& \frac{Q_{k-1} \left[\frac{1}{d_{k}} - \frac{1}{d_{u}}\right] + Q_{u}\left[\frac{1}{d_{k-1}} - \frac{1}{d_k}\right]}{\left[\frac{1}{d_{k-1}} - \frac{1}{d_{u}}\right]}. 
\eea
Now, note that from (\ref{eq:algo_lambda_2}), for every $n = k,\ldots,u-1$
\be
\label{eq:l6p_4}
\lambda_n^{i} =  \frac{Q_{k-1} \left[\frac{1}{d_{n}} - \frac{1}{d_{u}}\right] + Q_{u}\left[\frac{1}{d_{k-1}} - \frac{1}{d_n}\right]}{\left[\frac{1}{d_{k-1}} - \frac{1}{d_{u}}\right]}-Q_n.
\ee
From (\ref{eq:l6p_3}) and (\ref{eq:l6p_4}), clearly, $\lambda_k^{i} > 0$.
Now, from (\ref{eq:l6p_2}) and (\ref{eq:l6p_4}), it follows that for
every $n = k+1,\ldots,u-1$
\beas
\lambda_n^{i} - \lambda_n^{i-1} &=& \frac{Q_{k-1} \left[\frac{1}{d_{n}} - \frac{1}{d_{u}}\right] + Q_{u}\left[\frac{1}{d_{k-1}} - \frac{1}{d_n}\right]}{\left[\frac{1}{d_{k-1}} - \frac{1}{d_{u}}\right]} - \frac{Q_k \left[\frac{1}{d_n} - \frac{1}{d_u}\right] + Q_u\left[\frac{1}{d_k} - \frac{1}{d_n}\right]}{\left[\frac{1}{d_k} - \frac{1}{d_u}\right]} \\
&=& \frac{\left[\frac{1}{d_n} - \frac{1}{d_u}\right]}{\left[\frac{1}{d_k} - \frac{1}{d_u}\right]} \lambda_k^i  \ge 0.
\eeas
The last inequality follows from the fact that $d_n \le d_{n+1}$ for every $n$
and $ \lambda_k^i > 0$.
Furthermore, by  Lemma~\ref{lemma:change}, for $n\not\in\{k,\ldots,u-1\}$, $\lambda_n^{i-1} = \lambda_n^{i}$.
Thus, the result follows.
\endproof

\begin{lemma}
\label{lemma:l7}
Let index $k>1$ be added to the set $\mathcal{E}^{i-1}$ in the
$i^{\mathrm{th}}$ iteration. Also, let 
$\{v+1,\ldots,k-1\}\cup\{k+1,\ldots,u-1\} \subseteq \mathcal{E}^{i-1}$ 
and $\{v,u\} \subseteq \mathcal{A}^{i}$.
Then, $\lambda_n^{i} - \lambda_n^{i-1} \ge 0$ for every $n$.
\end{lemma}
\proof
Since index $k$ is added to $\mathcal{E}^{i-1}$ in the
$i^{\mathrm{th}}$ iteration, we know the following.
First, $\lambda_k^{i-1} = 0$. Second, from (\ref{eq:algo1})
\be
\label{eq:l7p_1}
Q_k < \frac{(Q_{k-1} + \lambda_{k-1}^{i-1})\left[\frac{1}{d_k} - \frac{1}{d_{k+1}}\right] + (Q_{k+1} + \lambda_{k+1}^{i-1}) )\left[\frac{1}{d_{k-1}} - \frac{1}{d_{k}}\right]}{\left[\frac{1}{d_{k-1}} - \frac{1}{d_{k+1}}\right]}.
\ee

Moreover, since 
$\{v+1,\ldots,k-1\}\cup\{k+1,\ldots,u-1\} \subseteq \mathcal{E}^{i-1}$
and $\{v,k,u\} \subseteq \mathcal{A}^{i-1}$, we also know that
for every $n \in \{v+1,\ldots,k-1\}$,
\bea
\label{eq:l7p_2}
\lambda_n^{i-1} = \frac{Q_v \left[\frac{1}{d_n} - \frac{1}{d_k}\right] + Q_k\left[\frac{1}{d_v} - \frac{1}{d_n}\right]}{\left[\frac{1}{d_v} - \frac{1}{d_k}\right]} - Q_n,
\eea
and for every $n \in \{k+1,\ldots,u-1\}$,
\bea
\label{eq:l7p_2.5}
\lambda_n^{i-1} = \frac{Q_k \left[\frac{1}{d_n} - \frac{1}{d_u}\right] + Q_u\left[\frac{1}{d_k} - \frac{1}{d_n}\right]}{\left[\frac{1}{d_k} - \frac{1}{d_u}\right]} - Q_n.
\eea
Now, substituting $\lambda_{k-1}^{i-1}$ and $\lambda_{k+1}^{i-1}$ from 
(\ref{eq:l7p_2}) and (\ref{eq:l7p_2.5}), respectively, in (\ref{eq:l7p_1}),
we obtain
\bea
Q_k &<& \frac{\frac{Q_v \left[\frac{1}{d_{k-1}} - \frac{1}{d_k}\right] + Q_k\left[\frac{1}{d_v} - \frac{1}{d_{k-1}}\right]}{\left[\frac{1}{d_v} - \frac{1}{d_k}\right]}\left[\frac{1}{d_{k}} - \frac{1}{d_{k+1}}\right] + \frac{Q_k \left[\frac{1}{d_{k+1}} - \frac{1}{d_u}\right] + Q_u\left[\frac{1}{d_k} - \frac{1}{d_{k+1}}\right]}{\left[\frac{1}{d_k} - \frac{1}{d_u}\right]}\left[\frac{1}{d_{k-1}} - \frac{1}{d_{k}}\right]}{\left[\frac{1}{d_{k-1}} - \frac{1}{d_{k+1}}\right]} \nonumber \\ 
\label{eq:l7p_3}
\Longrightarrow Q_k &<& \frac{Q_{v} \left[\frac{1}{d_{k}} - \frac{1}{d_{u}}\right] + Q_{u}\left[\frac{1}{d_{v}} - \frac{1}{d_k}\right]}{\left[\frac{1}{d_{v}} - \frac{1}{d_{u}}\right]}. 
\eea
Now, note that from (\ref{eq:algo_lambda_2}), for every $n = v+1,\ldots,u-1$
\be
\label{eq:l7p_4}
\lambda_n^{i} =  \frac{Q_{v} \left[\frac{1}{d_{n}} - \frac{1}{d_{u}}\right] + Q_{u}\left[\frac{1}{d_{v}} - \frac{1}{d_n}\right]}{\left[\frac{1}{d_{v}} - \frac{1}{d_{u}}\right]}-Q_n.
\ee
From (\ref{eq:l7p_3}) and (\ref{eq:l7p_4}), clearly, $\lambda_k^{i} > 0$.
Now, from (\ref{eq:l7p_2}) and (\ref{eq:l7p_4}), it follows that for
every $n = v+1,\ldots,k-1$
\beas
\lambda_n^{i} - \lambda_n^{i-1} &=& \frac{Q_{v} \left[\frac{1}{d_{n}} - \frac{1}{d_{u}}\right] + Q_{u}\left[\frac{1}{d_{v}} - \frac{1}{d_n}\right]}{\left[\frac{1}{d_{v}} - \frac{1}{d_{u}}\right]} - \frac{Q_v \left[\frac{1}{d_n} - \frac{1}{d_k}\right] + Q_k\left[\frac{1}{d_v} - \frac{1}{d_n}\right]}{\left[\frac{1}{d_v} - \frac{1}{d_k}\right]} \\
&=& \frac{\left[\frac{1}{d_v} - \frac{1}{d_n}\right]}{\left[\frac{1}{d_v} - \frac{1}{d_k}\right]} \lambda_k^i  \ge 0.
\eeas
The last inequality follows from the fact that $d_n \le d_{n+1}$ for every $n$
and $ \lambda_k^i > 0$.
Moreover, from (\ref{eq:l7p_2.5}) and (\ref{eq:l7p_4}), it follows that for
every $n = k+1,\ldots,u-1$
\beas
\lambda_n^{i} - \lambda_n^{i-1} &=& \frac{Q_{v} \left[\frac{1}{d_{n}} - \frac{1}{d_{u}}\right] + Q_{u}\left[\frac{1}{d_{v}} - \frac{1}{d_n}\right]}{\left[\frac{1}{d_{v}} - \frac{1}{d_{u}}\right]} - \frac{Q_k \left[\frac{1}{d_n} - \frac{1}{d_u}\right] + Q_u\left[\frac{1}{d_k} - \frac{1}{d_n}\right]}{\left[\frac{1}{d_k} - \frac{1}{d_u}\right]} \\
&=& \frac{\left[\frac{1}{d_n} - \frac{1}{d_u}\right]}{\left[\frac{1}{d_k} - \frac{1}{d_u}\right]} \lambda_k^i  \ge 0.
\eeas

Furthermore, by  Lemma~\ref{lemma:change}, for $n\not\in\{v+1,\ldots,u-1\}$, $\lambda_n^{i-1} = \lambda_n^{i}$.
Thus, the result follows.
\endproof

\begin{lemma}
\label{lemma:l8}
Let index $k>1$ be added to the set $\mathcal{E}^{i-1}$ in the
$i^{\mathrm{th}}$ iteration. Also, let 
$\{1,\ldots,k-1\}\cup\{k+1,\ldots,u-1\} \subseteq \mathcal{E}^{i-1}$ 
and $u \in \mathcal{A}^{i}$.
Then, $\lambda_n^{i} - \lambda_n^{i-1} \ge 0$ for every $n$.
\end{lemma}
\proof
Since index $k$ is added to $\mathcal{E}^{i-1}$ in the
$i^{\mathrm{th}}$ iteration, we know the following.
First, $\lambda_k^{i-1} = 0$. Second, from (\ref{eq:algo1})
\be
\label{eq:l8p_1}
Q_k < \frac{(Q_{k-1} + \lambda_{k-1}^{i-1})\left[\frac{1}{d_k} - \frac{1}{d_{k+1}}\right] + (Q_{k+1} + \lambda_{k+1}^{i-1}) )\left[\frac{1}{d_{k-1}} - \frac{1}{d_{k}}\right]}{\left[\frac{1}{d_{k-1}} - \frac{1}{d_{k+1}}\right]}.
\ee

Moreover, since 
$\{1,\ldots,k-1\}\cup\{k+1,\ldots,u-1\} \subseteq \mathcal{E}^{i-1}$
and $\{k,u\} \subseteq \mathcal{A}^{i-1}$, we also know that
for every $n \in \{1,\ldots,k-1\}$,
\be
\label{eq:l8p_2}
\lambda_n^{i-1} = V N_0 \left[\frac{1}{d_n} - \frac{1}{d_k}\right] +(Q_k-Q_n),
\ee
and for every $n \in \{k+1,\ldots,u-1\}$,
\be
\label{eq:l8p_2.5}
\lambda_n^{i-1} = \frac{Q_k \left[\frac{1}{d_n} - \frac{1}{d_u}\right] + Q_u\left[\frac{1}{d_k} - \frac{1}{d_n}\right]}{\left[\frac{1}{d_k} - \frac{1}{d_u}\right]} - Q_n.
\ee
Now, substituting $\lambda_{k-1}^{i-1}$ and $\lambda_{k+1}^{i-1}$ from 
(\ref{eq:l8p_2}) and (\ref{eq:l8p_2.5}), respectively, in (\ref{eq:l8p_1}),
we obtain
\bea
Q_k &<& \frac{\left[V N_0 \left[\frac{1}{d_{k-1}} - \frac{1}{d_k}\right] +Q_k\right]\left[\frac{1}{d_{k}} - \frac{1}{d_{k+1}}\right] + \frac{Q_k \left[\frac{1}{d_{k+1}} - \frac{1}{d_u}\right] + Q_u\left[\frac{1}{d_k} - \frac{1}{d_{k+1}}\right]}{\left[\frac{1}{d_k} - \frac{1}{d_u}\right]}\left[\frac{1}{d_{k-1}} - \frac{1}{d_{k}}\right]}{\left[\frac{1}{d_{k-1}} - \frac{1}{d_{k+1}}\right]} \nonumber \\ 
\label{eq:l8p_3}
\Longrightarrow Q_k &<& V N_0 \left[\frac{1}{d_{k}} - \frac{1}{d_u}\right] +Q_u.
\eea
Now, note that from (\ref{eq:algo_lambda_2}), for every $n = 1,\ldots,u-1$
\be
\label{eq:l8p_4}
\lambda_n^{i} =  V N_0 \left[\frac{1}{d_{n}} - \frac{1}{d_u}\right] + (Q_u-Q_n).
\ee
From (\ref{eq:l8p_3}) and (\ref{eq:l8p_4}), clearly, $\lambda_k^{i} > 0$.
Now, from (\ref{eq:l8p_2}) and (\ref{eq:l8p_4}), it follows that for
every $n = 1,\ldots,k-1$
\beas
\lambda_n^{i} - \lambda_n^{i-1} &=& V N_0 \left[\frac{1}{d_{n}} - \frac{1}{d_u}\right] + (Q_u-Q_n) - V N_0 \left[\frac{1}{d_{n}} - \frac{1}{d_k}\right] - (Q_k-Q_n) \\
&=& \lambda_k^i  \ge 0.
\eeas
The last inequality follows from the fact that
and $ \lambda_k^i > 0$.
Moreover, from (\ref{eq:l8p_2.5}) and (\ref{eq:l8p_4}), it follows that for
every $n = k+1,\ldots,u-1$
\beas
\lambda_n^{i} - \lambda_n^{i-1} &=& V N_0 \left[\frac{1}{d_{n}} - \frac{1}{d_u}\right] + Q_u - \frac{Q_k \left[\frac{1}{d_n} - \frac{1}{d_u}\right] + Q_u\left[\frac{1}{d_k} - \frac{1}{d_n}\right]}{\left[\frac{1}{d_k} - \frac{1}{d_u}\right]} \\
&=& \left[\frac{1}{d_n} - \frac{1}{d_u}\right] \lambda_k^i  \ge 0.
\eeas

Furthermore, by  Lemma~\ref{lemma:change}, for $n\not\in\{1,\ldots,u-1\}$, $\lambda_n^{i-1} = \lambda_n^{i}$.
Thus, the result follows.
\endproof

\end{document}